\documentclass[journal]{IEEEtran}
 \usepackage{tabularx}
\usepackage{multirow}
\usepackage{amsmath,amssymb}

\usepackage{changepage}

\usepackage[utf8x]{inputenc}

\usepackage{textcomp,marvosym}

\usepackage{cite}

\usepackage{nameref,hyperref}

\usepackage{microtype}
\DisableLigatures[f]{encoding = *, family = * }

\usepackage[table]{xcolor}

\usepackage{array}

\newcolumntype{+}{!{\vrule width 2pt}}

\newlength\savedwidth

\makeatletter
\renewcommand{\@biblabel}[1]{\quad#1.}
\makeatother

\usepackage{lastpage,fancyhdr,graphicx}

\title{A temporal-to-spatial deep convolutional neural network for classification of hand movements from multichannel electromyography data}

\author{Adam~Hartwell, Viskan~Kadirkamanathan and~Sean~R.~Anderson%
\thanks{
Adam~Hartwell, Viskan~Kadirkamanathan and Sean~R.~Anderson {\tt\small \{a.hartwell,visakan,s.anderson\} @sheffield.ac.uk} are with the Department of Automatic Control and Systems Engineering, The University of Sheffield, Sheffield, S1 3JD, UK
}
}

\begin{document}

\maketitle

\begin{abstract}
    Deep convolutional neural networks (CNNs) are appealing for the purpose of classification of hand movements from surface electromyography (sEMG) data because they have the ability to perform automated person-specific feature extraction from raw data. 
    In this paper, we make the novel contribution of proposing and evaluating a design for the early processing layers in the deep CNN for multichannel sEMG. Specifically, we propose a novel temporal-to-spatial (TtS) CNN architecture, where the first layer performs convolution separately on each sEMG channel to extract temporal features. This is motivated by the idea that sEMG signals in each channel are mediated by one or a small subset of muscles, whose temporal activation patterns are associated with the signature features of a gesture. The temporal layer captures these signature features for each channel separately, which are then spatially mixed in successive layers to recognise a specific gesture. 
    A practical advantage is that this approach also makes the CNN simple to design for different sample rates. 
    We use NinaPro database 1 (27 subjects and 52 movements + rest), sampled at 100 Hz, and database 2 (40 subjects and 40 movements + rest), sampled at 2 kHz, to evaluate our proposed CNN design. We benchmark against a feature-based support vector machine (SVM) classifier, two CNNs from the literature, and an additional standard design of CNN. We find that our novel TtS CNN design achieves 66.6\% per-class accuracy on database 1, and 67.8\% on database 2, and that the TtS CNN outperforms all other compared classifiers using a statistical hypothesis test at the $2\%$ significance level. 
\end{abstract}



\section{Introduction}
    Machine learning is an essential tool for extracting user intention for control of devices. For hand movement recognition \cite{Mitra2007}, this can be done using bioelectric signals \cite{AsghariOskoei2007, Farina2014}, ultrasound \cite{Huang2017}, cameras \cite{Patsadu2012} or motion capture using smart gloves \cite{Ju2009}. Hand movement classification from surface electromyography (sEMG) has been performed using various methods, such as linear discriminant analysis \cite{Duan2018}, support vector machines (SVMs) \cite{Atzori2014d, Castellini2009, Quitadamo2017}, neural networks \cite{Atzori2014d, Castellini2009, Duan2016}, neurofuzzy \cite{Balbinot2013, Khezri2011} and mixtures of experts \cite{Baldacchino2018}. These conventional classifiers have mostly been applied to small numbers of movement classes:  e.g. 5-7 movements \cite{Castellini2009, Shin2014, Balbinot2013, Khezri2011, Khushaba2009, Khezri2007, Huang2005, Lucas2008a} or 9-12 movements \cite{Ju2013, Khushaba2007, Tenore2009, Zhou2010, Ortiz-Catalan2012}. Typically, for these conventional classifiers, feature extraction is performed using fixed features e.g. wavelets or Fourier transforms \cite{Canal2010, Reaz2006}. This approach to feature extraction is difficult and limited because the features have to be carefully engineered by domain experts and are not person-specific. 
    
    

    Deep convolutional neural networks (CNNs) \cite{LeCun2015} have the potential to improve and simplify sEMG classifier systems, due to their ability to perform automated person-specific feature extraction from raw data inputs. Previous work has already shown some potential for CNNs in this domain: an early study in comparison to support vector machines (SVMs) has demonstrated that CNNs can be competitive although not necessarily outperform feature-based classifiers \cite{Atzori2016}; subsequent studies have shown that CNNs can outperform SVMs in the context of re-calibration \cite{Zhai2017} and regression for motor control \cite{Ameri2019}; research on classifying sEMG signals from instantaneous measurements has shown promise for low-latency systems \cite{Geng2016}; experiments across multiple days have shown improvements in CNNs compared to linear discriminant analysis \cite{ZiaurRehman2018}.  However, no particular CNN designs stand-out as optimal for sEMG classifiers, and there are no particular guidelines that a user can follow, and so optimal CNN design for sEMG is still an open question and one that requires study.  

    The focus of this paper is the design of a novel CNN architecture where the lowest layers of the network  perform convolution only along the temporal direction of each separate sEMG channel in order to extract temporal features. This is motivated by the idea that sEMG signals in each channel are mediated by one or a small subset of muscles, whose temporal activation patterns are associated with the signature features of a gesture. The temporal CNN layer captures these signature features for each channel separately, which are then spatially mixed in successive CNN layers to recognise a specific gesture. We label this CNN design a Temporal-to-Spatial CNN (TtS CNN). 
    
    The TtS CNN has some similarity to the wide and successful use of temporal feature extraction from multi-channel sEMG in conventional feature-based classifiers, e.g. using Fourier and wavelet transforms, which operate on each sEMG channel separately \cite{Phinyomark2013}. 
    A practical advantage of the TtS CNN is that the network architecture can be easily re-designed for different input data sizes, caused by differences in sampling rate or window length between sEMG classifier systems. 
    The low-level temporal convolutions are also combined here for the first time with a modified version of the compression techniques used in SqueezeNet - a technique that tends to reduce network size for a given performance level \cite{Iandola2016}.

    
    To evaluate the TtS CNN we use NinaPro databases 1 and 2 \cite{Atzori, Atzori2014a, NinaProProject2015}, which are an open-source collection of sEMG data associated with many hand movements, where in database~1 there are 27 subjects performing 53 movements (52 + \textit{rest}), and in database~2 there are 40 subjects performing 41 movements (40 + \textit{rest}). Ninapro database 1 consists of multi-channel sEMG sampled at 100 Hz, whilst Ninapro database 2 is sampled at 2 kHz. These databases are ideal for demonstrating the advantage of the temporal convolution layer in our proposed CNN design in comparison to other CNNs, when having to re-design the CNN for different sample rates. 
    
    We compare and benchmark the TtS CNN design against the CNNs from Atzori \textit{et al} \cite{Atzori2016}, Geng \textit{et al} \cite{Geng2016, Du2017} and our own generic CNN design (without the temporal convolution layer). We also compare to a feature-based classification method in the form of a Support Vector Machine (SVM) using the following features: marginal Discrete Wavelet Transform \cite{Lucas2008}, Mean Absolute Value (MAV) \cite{Hudgins} and Waveform Length (WL) \cite{Hudgins}. We use robust validation methods (stratified multi-split cross-validation) to evaluate the performance of each classifier and a per-class method of measuring accuracy that is resistant to bias caused by data imbalance - the macro-average accuracy \cite{Sokolova2009, Ortiz-Catalan2015}. We also perform a statistical comparison of classifiers based on a specialist method for multiple data sets (because each classifier is trained separately on each human subject) \cite{Dem2006}, in order to demonstrate a significant performance improvement over the comparison classifiers. 
    
    

\section{Methodology} \label{Methods} 
    \subsection{Standard Convolutional Neural Network Design}
        In this section we give a brief overview and background on CNN design. All specific details are for the networks we designed; reproduction of previous works used implementation specifics defined in their respective papers.
    
        The input to our classifiers is a window of sEMG data, $X\in\mathbb{R}^{n_s\times n_c}$, where $n_s$ is the number of samples and $n_c$ the number of sEMG channels. The main building block of CNNs is the convolutional layer, where a 2D convolution is a single 2D map, indexed by $k$, in layer $l$, is $Z^{(l,k)} \in \mathbb{R}^{r_l \times c_l}$, where $Z^{0,1}=X$. At each layer there is a stack of $d_l$ maps, i.e. a 3D volume of dimension $r_l\times c_l \times d_l$. The value of a unit, $z_{r,c}^{(l,k)}$, at location $(r,c)$, in the map $Z^{(l,k)}$, is given by
        \begin{equation} 
             z_{r,c}^{(l,k)} = h_a \left( \left( \sum_{m=1}^{d_{l-1}}  \sum_{i=1}^{R_l} \sum_{j=1}^{C_l} w_{i,j}^{(l,k,m)} z^{(l-1,m)}_{\tilde{r} + i, \tilde{c} + j} \right) + b^{(l,k)}   \right) 
        \end{equation}
        where $z_{r,c}^{(l,k)}$ is the neuron output at location $(r,c)$, for $r=1,\ldots,r_l$, $c=1,\ldots,c_l$, $R_l \times C_l$ is the convolution filter size, the convolution filter indexed by $k$, for $k=1,\ldots,d_l$, is composed of the adjustable CNN weights $w_{i,j}^{(l,k,m)}$, $b^{(l,k)}$ is a bias term, and $\tilde{r}=r - \lceil R_l/2 \rceil$ and $\tilde{c}=c - \lceil C_l/2 \rceil$ for odd valued $R_l$ and $C_l$. $h_{a}(.)$ is the activation function of the neuron, defined here, for all but the final layer of each network, as the leaky rectified linear unit (LReLU) \cite{maas2013rectifier, Xu2015}, where
        \begin{equation} 
            h_{a}(x) = 
            \begin{cases}
                x, \ \ \ x \geq 0\\
                \alpha x,\ x < 0\\
            \end{cases}
            \label{lrelu}
        \end{equation}
        where $0 < \alpha < 1$.
        
        The final layer, which performs the classification, is a dense layer with a softmax activation function in each case, which is defined as 
        \begin{equation} 
             z^{*}_j = \exp \left( \tilde{z}_j^{(l)} \right) \times \left( \sum^{M - 1}_{k = 0} \exp \left( \tilde{z}_k^{(l)} \right) \right)^{-1} 
        \end{equation}
        for class $j=1,\ldots,M-1$, where $M$ is the number of classes,  $z^{*}_j$ is the normalised output of the softmax layer for class $j$ and
        \begin{equation}
            \tilde{z}_r^{(l)} = \left( \sum_{m=1}^{d_{l-1}} \sum_{j=1}^{r_{l-1}} \sum_{i=1}^{c_{l-1}} w_{i,j}^{(l,m)} z^{(l-1,m)}_{i,j} \right) + b^{(l,r)}
        \end{equation}
        for $r=1,\ldots,M-1$.
        
        The network weights were trained using the cross-entropy loss function, for $N$ data samples and $M$ classes,
         \begin{equation} 
            L(\Theta) = - \sum_{i=1}^{N} \sum_{j=0}^{M-1} 1 \lbrace y^{(i)} = j \rbrace \log z^{*}_{ij} 
        \end{equation}
        where $\Theta$ is the set of all CNN parameters, including weights and biases from all layers, $z^{*}_{ij}$ is the softmax output for data sample $i$ and prediction of class $j$, $y^{(i)}$ is the true class label for data sample $i$,  $1\{.\}$ is the indicator function, i.e. $1\{.\}=1$ for true and $1\{.\}=0$ for false. The weight parameters were randomly initialised using the Glorot uniform kernel \cite{Glorot2010}, and bias parameters were initialised to zero.
        
        During training a class weighting prior was also used to aid training based on the frequency of examples in the training set
        \begin{equation}  \label{eq:class_weight}
            \gamma_{j} = 1 + \log_{2}{\frac{n_{max}}{n_j}}
        \end{equation}
        where $\gamma_{j}$ is the weighting of movement $j$, $n_{max}$ is the number of examples of the most represented movement and $n_j$ is the number of examples of movement $j$.
        
        We implemented our work in Python primarily using Keras \cite{chollet2015keras} with Tensorflow \cite{Abadi2016} except for re-implementation of the Geng \textit{et al} network \cite{Geng2016} which was re-implemented in MXNet \cite{Chen2015}. Training was performed using an NVIDIA Tesla K40 GPU with 12 GB RAM. 
        
        Decisions about hyper-parameters were made based on a rapid-prototyping approach that evaluated potential modifications on a small subset of the data consisting of one validation fold of three subjects worth of data. The subjects and validation split were generated randomly for each different parameter that was worked on. 
        
        The selection for hyper-parameters to test was driven by best practice for CNNs in other domains, sEMG domain knowledge and informed by previous prototyping results.
        
    \subsection{Temporal-to-Spatial Network Design} \label{sec:design}
\begin{figure}
            \centering
            \includegraphics[width=1\columnwidth,keepaspectratio]{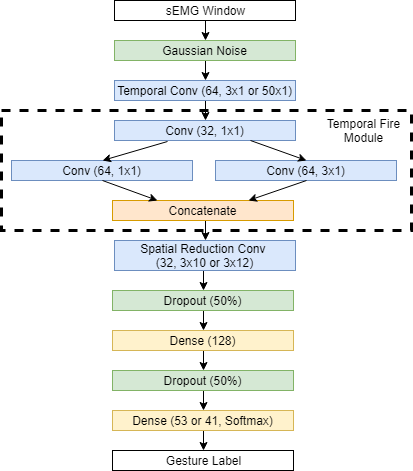}
            \caption{Graphical representation of the TtS network. Brackets show number of filters followed by filter size. The X or Y on filter sizes shows the size for database~1 and database~2 respectively.}
            \label{fig:networkDia}
        \end{figure}

        Tables \ref{tab:network_db1} and \ref{tab:network_db2} show our main network designs for the two databases. Fig \ref{fig:networkDia} shows a graphical representation.
        
        \begin{table*}
        \scriptsize
            \centering
            \caption{Breakdown of TtS network architecture for database~1. Channels last format.}
            \begin{tabular}{cccccc|c|}
            \hline
            \rowcolor[HTML]{6665CD} 
            \multicolumn{1}{|c|}{\cellcolor[HTML]{6665CD}\textbf{Layer Type}} & \multicolumn{1}{c|}{\cellcolor[HTML]{6665CD}\textbf{Output Size}} & \multicolumn{1}{c|}{\cellcolor[HTML]{6665CD}\textbf{\# Filters}} & \multicolumn{1}{c|}{\cellcolor[HTML]{6665CD}\textbf{Filter Size}} & \multicolumn{1}{c|}{\cellcolor[HTML]{6665CD}\textbf{Stride}} & 
            \textbf{Activation}      &  
            \textbf{\# Params}         \\ \hline
            \multicolumn{1}{|c|}{Input EMG}                                   & \multicolumn{1}{c|}{15x10x1}                                      & 
            \multicolumn{1}{c|}{}  & 
            \multicolumn{1}{c|}{}   & 
            \multicolumn{1}{c|}{}                                        &   
            
                                     &                                  \\ \hline
            \rowcolor[HTML]{BBDAFF} 
            \multicolumn{1}{|c|}{\cellcolor[HTML]{BBDAFF}Gaussian Noise}      & \multicolumn{1}{c|}{\cellcolor[HTML]{BBDAFF}15x10x1}              & \multicolumn{1}{c|}{\cellcolor[HTML]{BBDAFF}}                    & \multicolumn{1}{c|}{\cellcolor[HTML]{BBDAFF}}                     & \multicolumn{1}{c|}{\cellcolor[HTML]{BBDAFF}}                &  
             $\alpha = 0.001$           &                                   \\ \hline
            \multicolumn{1}{|c|}{Conv (Temporal)}                             & \multicolumn{1}{c|}{15x10x64}                                     & 
            \multicolumn{1}{c|}{64}                                          & 
            \multicolumn{1}{c|}{3x1}                                          & \multicolumn{1}{c|}{1x1}                                     & 
             LReLU     &            256                             \\ \hline
            \rowcolor[HTML]{BBDAFF} 
            \multicolumn{1}{|c|}{\cellcolor[HTML]{BBDAFF}Temporal Fire}       & \multicolumn{1}{c|}{\cellcolor[HTML]{BBDAFF}15x10x128}            & \multicolumn{1}{c|}{\cellcolor[HTML]{BBDAFF}(32, 64, 64)}        & \multicolumn{1}{c|}{\cellcolor[HTML]{BBDAFF}(1x1, 1x1, 3x1)}      & \multicolumn{1}{c|}{\cellcolor[HTML]{BBDAFF}1x1}             & 
             LReLU      &                10,400               \\ \hline
            \multicolumn{1}{|c|}{Conv (Spatial)}                              & \multicolumn{1}{c|}{15x10x32}                                     & 
            \multicolumn{1}{c|}{32}                                          & \multicolumn{1}{c|}{3x10}                                         & \multicolumn{1}{c|}{1x1}                                     & 
             LReLU        &              122,912                         \\ \hline
            \rowcolor[HTML]{BBDAFF} 
            \multicolumn{1}{|c|}{\cellcolor[HTML]{BBDAFF}Dropout}             & \multicolumn{1}{c|}{\cellcolor[HTML]{BBDAFF}4,800}                & \multicolumn{1}{c|}{\cellcolor[HTML]{BBDAFF}}                    & \multicolumn{1}{c|}{\cellcolor[HTML]{BBDAFF}}                     & \multicolumn{1}{c|}{\cellcolor[HTML]{BBDAFF}}                &    
             rate = 0.5                   &                               \\ \hline
            \multicolumn{1}{|c|}{Dense}                                       & \multicolumn{1}{c|}{128}                                          & \multicolumn{1}{c|}{128}                                         & 
            \multicolumn{1}{c|}{}                                             & 
            \multicolumn{1}{c|}{}                                        &   
             LReLU       &          614,528                         \\ \hline
            \rowcolor[HTML]{BBDAFF} 
            \multicolumn{1}{|c|}{\cellcolor[HTML]{BBDAFF}Dropout}             & \multicolumn{1}{c|}{\cellcolor[HTML]{BBDAFF}128}                  & \multicolumn{1}{c|}{\cellcolor[HTML]{BBDAFF}}                    & \multicolumn{1}{c|}{\cellcolor[HTML]{BBDAFF}}                     & \multicolumn{1}{c|}{\cellcolor[HTML]{BBDAFF}}                &  
             rate = 0.5                   &                        \\ \hline
            \multicolumn{1}{|c|}{Dense}                                       & 
            \multicolumn{1}{c|}{53}                                           & 
            \multicolumn{1}{c|}{53}                                          & 
            \multicolumn{1}{c|}{}                                             & 
            \multicolumn{1}{c|}{}                                        &  
             Softmax                      & 6,837                           \\ \hline
            \multicolumn{1}{l}{}                                              & 
            \multicolumn{1}{l}{}                                              & 
            \multicolumn{1}{l}{}                                             & 
            \multicolumn{1}{l}{}                                              & 
            \multicolumn{1}{l}{}                                         & 
             \cellcolor[HTML]{BBDAFF}\textbf{Total} &  \cellcolor[HTML]{BBDAFF}754,933 \\ \cline{6-7} 
            \end{tabular}
            
            \label{tab:network_db1}
        \end{table*}

         \begin{table*}
            \centering
            \scriptsize
            \caption{Breakdown of TtS network architecture for database~2. Channels last format.}
            \begin{tabular}{cccccc|c|}
            \hline
            \rowcolor[HTML]{6665CD} 
            \multicolumn{1}{|c|}{\cellcolor[HTML]{6665CD}\textbf{Layer Type}} & \multicolumn{1}{c|}{\cellcolor[HTML]{6665CD}\textbf{Output Size}} & \multicolumn{1}{c|}{\cellcolor[HTML]{6665CD}\textbf{\# Filters}} & \multicolumn{1}{c|}{\cellcolor[HTML]{6665CD}\textbf{Filter Size}} & \multicolumn{1}{c|}{\cellcolor[HTML]{6665CD}\textbf{Stride}} &  \textbf{Activation}    &  \textbf{\# Params}          \\ \hline
            \multicolumn{1}{|c|}{Input EMG}                                   & \multicolumn{1}{c|}{300x12x1}                                     & \multicolumn{1}{c|}{}                                            & \multicolumn{1}{c|}{}                                             & \multicolumn{1}{c|}{}                                        &                             &                                             \\ \hline
            \rowcolor[HTML]{BBDAFF} 
            \multicolumn{1}{|c|}{\cellcolor[HTML]{BBDAFF}Gaussian Noise}      & \multicolumn{1}{c|}{\cellcolor[HTML]{BBDAFF}300x12x1}             & \multicolumn{1}{c|}{\cellcolor[HTML]{BBDAFF}}                    & \multicolumn{1}{c|}{\cellcolor[HTML]{BBDAFF}}                     & \multicolumn{1}{c|}{\cellcolor[HTML]{BBDAFF}}                &    $\alpha = 0.001$         &                           \\ \hline
            \multicolumn{1}{|c|}{Conv (Temporal)}                             & \multicolumn{1}{c|}{12x12x64}                                     & \multicolumn{1}{c|}{64}                                          & \multicolumn{1}{c|}{50x1}                                         & \multicolumn{1}{c|}{25x1}                                    &  LReLU      &               3,264                        \\ \hline
            \rowcolor[HTML]{BBDAFF} 
            \multicolumn{1}{|c|}{\cellcolor[HTML]{BBDAFF}Temporal Fire}       & \multicolumn{1}{c|}{\cellcolor[HTML]{BBDAFF}12x12x128}            & \multicolumn{1}{c|}{\cellcolor[HTML]{BBDAFF}(32, 64, 64)}        & \multicolumn{1}{c|}{\cellcolor[HTML]{BBDAFF}(1x1, 1x1, 3x1)}      & \multicolumn{1}{c|}{\cellcolor[HTML]{BBDAFF}1x1}             &  LReLU      &                10,400                        \\ \hline
            \multicolumn{1}{|c|}{Conv (Spatial)}                              & \multicolumn{1}{c|}{12x12x32}                                     & \multicolumn{1}{c|}{32}                                          & \multicolumn{1}{c|}{3x12}                                         & \multicolumn{1}{c|}{1x1}                                     &  LReLU     &                147,488                      \\ \hline
            \rowcolor[HTML]{BBDAFF} 
            \multicolumn{1}{|c|}{\cellcolor[HTML]{BBDAFF}Dropout}             & \multicolumn{1}{c|}{\cellcolor[HTML]{BBDAFF}4,608}                & \multicolumn{1}{c|}{\cellcolor[HTML]{BBDAFF}}                    & \multicolumn{1}{c|}{\cellcolor[HTML]{BBDAFF}}                     & \multicolumn{1}{c|}{\cellcolor[HTML]{BBDAFF}}                &     rate = 0.5               &                              \\ \hline
            \multicolumn{1}{|c|}{Dense}                                       & \multicolumn{1}{c|}{128}                                          & \multicolumn{1}{c|}{128}                                         & \multicolumn{1}{c|}{}                                             & \multicolumn{1}{c|}{}                                        &     LReLU    &                  589,952                        \\ \hline
            \rowcolor[HTML]{BBDAFF} 
            \multicolumn{1}{|c|}{\cellcolor[HTML]{BBDAFF}Dropout}             & \multicolumn{1}{c|}{\cellcolor[HTML]{BBDAFF}128}                  & \multicolumn{1}{c|}{\cellcolor[HTML]{BBDAFF}}                    & \multicolumn{1}{c|}{\cellcolor[HTML]{BBDAFF}}                     & \multicolumn{1}{c|}{\cellcolor[HTML]{BBDAFF}}                &    rate = 0.5             &                            \\ \hline
            \multicolumn{1}{|c|}{Dense}                                       & \multicolumn{1}{c|}{41}                                           & \multicolumn{1}{c|}{41}                                          & \multicolumn{1}{c|}{}                                             & \multicolumn{1}{c|}{}                                        &     Softmax               &                  5,289                        \\ \hline
            \multicolumn{1}{l}{}                                              & \multicolumn{1}{l}{}                                              & \multicolumn{1}{l}{}                                             & \multicolumn{1}{l}{}                                              & \multicolumn{1}{l}{}                                         &  \cellcolor[HTML]{BBDAFF}\textbf{Total} & \cellcolor[HTML]{BBDAFF}756,393 \\ 
            \cline{6-7} 
            \end{tabular}

            \label{tab:network_db2}
        \end{table*}

        The key contributions of this paper lie in these network designs which encode domain knowledge into the network architecture. We shall call the network which incorporates all of the points below, the Temporal-to-Spatial (TtS) network based on how its architecture manipulates data flow.
        
        The convolutional layers (marked ``Conv (Temporal)" and ``Temporal Fire" in the tables) allow rapid non-linear expansion of input data and learning of complex low level features which we constrain in the lower layers to the temporal direction by using filters of size $N \times 1$ which perform convolution on only a single channel of sEMG data. This specifically encodes that we expect useful low level features to be temporal in nature i.e. not calculated across channels. This has shown to be the case with the majority of hand designed features which are calculated on a per-channel basis rather than across all channels \cite{Reaz2006, Scheme2014, Phinyomark2013}. 
        
        By enforcing the temporal constraint we greatly increase the likelihood of learning useful, generalisable features. Then by using large numbers of early filters we ensure complex expansions are possible that can tailor the temporal feature extraction to a specific subject. For instance this may be likened to selection of the best wavelet for a set of subjects when using the Discrete Wavelet Transform (DWT) as a hand crafted feature except we may now automatically tailor that to a specific subject and not be constrained by a pool of wavelets to select from.
        
        We also augment the Fire Module described in SqueezeNet \cite{Iandola2016}. The Fire Module consists of a 1x1 convolution whose output is fed to another 1x1 convolution and a 3x3 convolution which are then concatenated together to form the output. The aim of this design was model compression: allowing a network with far fewer parameters to compete with much larger networks in terms of performance. We reverse that idea here and modify the Fire Module to include temporal enforcement in order to boost performance. The key insight is that the Fire Module learns inter-filter connections i.e. connections between different features. Combinations of features have been shown to improve performance in traditional classification solutions to sEMG classification \cite{Pizzolato2017} and thus by encoding specifically that these are likely to be important we can further enhance performance
        
        To enforce temporal features we modify the 3x3 convolution in the Fire Module to be a 3x1 and rename it a ``Temporal Fire Module'' for clarity.
        
        This temporal first approach is diametrically opposite to Atzori \textit{et al} \cite{Atzori2016} and Geng \textit{et al} \cite{Geng2016, Du2017} who both explicitly perform spatial convolution first in their networks.
        
        The final convolutional layer (marked ``Spatial" in the tables) is the only convolution that allows inter-channel features. Intuitively, in a similar way to low level temporal feature enforcement, this layer encodes the idea that high level features will relate to combinations across channels e.g. patterns of channel activation relate to specific kinds of movements.
        
        The Gaussian noise layer helps prevent overfitting by introducing noise to incoming training samples. The $\alpha$ parameter was selected based on Atzori \textit{et al}'s \cite{Atzori2016} work. 
        
        Another key contribution, seen in the TtS and Baseline CNN designs, is adaption to different input sizes caused either by different window lengths or sampling frequencies. This is illustrated here in the implementation differences of the TtS network between databases~1 and 2 caused by the 20x higher sampling frequency in database~2 (Tables \ref{tab:network_db1} and \ref{tab:network_db2}). The key insight is that the first convolution can be expanded along the temporal direction to cover a similar span of time and the stride increased to also cover a similar time increment through the window. This approach drastically reduces the necessary parameters, compared to a direct conversion by updating the input shape, while still maintaining performance.
        
        For training we used the Adam algorithm \cite{Kingma2015} with a learning rate of $0.001$, $\beta_1 = 0.9$ and $\beta_2 = 0.999$. From our prototyping we found that 10 epochs was a sufficient training length for all cases except the baseline CNN on database~2 which only required 5.These choices as well as the other architectural and hyper-parameter choices not mentioned explicitly were identified by a combination of random search and manual tuning \cite{LeCun2015} cross validated using the training data from 10 randomly selected subjects from the main benchmark to ensure informational separation. 
    
    \subsection{Baseline CNN Design}

        \begin{table*}
            \centering
             \scriptsize
            \caption{Baseline CNN architecture for database~1. Channels last format.}
            \begin{tabular}{cccccc|c|}
            \hline
            \rowcolor[HTML]{6665CD} 
            \multicolumn{1}{|c|}{\cellcolor[HTML]{6665CD}\textbf{Layer Type}} & \multicolumn{1}{c|}{\cellcolor[HTML]{6665CD}\textbf{Output Size}} & \multicolumn{1}{c|}{\cellcolor[HTML]{6665CD}\textbf{\# Filters}} & \multicolumn{1}{c|}{\cellcolor[HTML]{6665CD}\textbf{Filter Size}} & \multicolumn{1}{c|}{\cellcolor[HTML]{6665CD}\textbf{Stride}} & \textbf{Activation}     &  \textbf{\# Params}          \\ \hline
            \multicolumn{1}{|c|}{Input EMG}                                   & \multicolumn{1}{c|}{15x10x1}                                      & \multicolumn{1}{c|}{}                                            & \multicolumn{1}{c|}{}                                             & \multicolumn{1}{c|}{}                                        &                                         &                                 \\ \hline
            \rowcolor[HTML]{BBDAFF} 
            \multicolumn{1}{|c|}{\cellcolor[HTML]{BBDAFF}Gaussian Noise}      & \multicolumn{1}{c|}{\cellcolor[HTML]{BBDAFF}15x10x1}              & \multicolumn{1}{c|}{\cellcolor[HTML]{BBDAFF}}                    & \multicolumn{1}{c|}{\cellcolor[HTML]{BBDAFF}}                     & \multicolumn{1}{c|}{\cellcolor[HTML]{BBDAFF}}                &    $\alpha = 0.001$        &                      \\ \hline
            \multicolumn{1}{|c|}{Conv}                                        & \multicolumn{1}{c|}{15x10x128}                                    & \multicolumn{1}{c|}{128}                                         & \multicolumn{1}{c|}{3x3}                                          & \multicolumn{1}{c|}{1x1}                                     &   LReLU    &  1,280               \\ \hline
            \rowcolor[HTML]{BBDAFF} 
            \multicolumn{1}{|c|}{\cellcolor[HTML]{BBDAFF}Conv}                & \multicolumn{1}{c|}{\cellcolor[HTML]{BBDAFF}15x10x64}             & \multicolumn{1}{c|}{\cellcolor[HTML]{BBDAFF}64}                  & \multicolumn{1}{c|}{\cellcolor[HTML]{BBDAFF}5x3}                  & \multicolumn{1}{c|}{\cellcolor[HTML]{BBDAFF}1x1}             &  LReLU    &  122,944              \\ \hline
            \multicolumn{1}{|c|}{Conv}                                        & \multicolumn{1}{c|}{15x10x32}                                     & \multicolumn{1}{c|}{32}                                          & \multicolumn{1}{c|}{5x3}                                          & \multicolumn{1}{c|}{1x1}                                     &   LReLU    &  30,752           \\ \hline
            \rowcolor[HTML]{BBDAFF} 
            \multicolumn{1}{|c|}{\cellcolor[HTML]{BBDAFF}Dropout}             & \multicolumn{1}{c|}{\cellcolor[HTML]{BBDAFF}4,800}                & \multicolumn{1}{c|}{\cellcolor[HTML]{BBDAFF}}                    & \multicolumn{1}{c|}{\cellcolor[HTML]{BBDAFF}}                     & \multicolumn{1}{c|}{\cellcolor[HTML]{BBDAFF}}                &    rate = 0.5               &                    \\ \hline
            \multicolumn{1}{|c|}{Dense}                                       & \multicolumn{1}{c|}{128}                                          & \multicolumn{1}{c|}{128}                                         & \multicolumn{1}{c|}{}                                             & \multicolumn{1}{c|}{}                                        &     LReLU    &    614,528          \\ \hline
            \rowcolor[HTML]{BBDAFF} 
            \multicolumn{1}{|c|}{\cellcolor[HTML]{BBDAFF}Dropout}             & \multicolumn{1}{c|}{\cellcolor[HTML]{BBDAFF}128}                  & \multicolumn{1}{c|}{\cellcolor[HTML]{BBDAFF}}                    & \multicolumn{1}{c|}{\cellcolor[HTML]{BBDAFF}}                     & \multicolumn{1}{c|}{\cellcolor[HTML]{BBDAFF}}                &    rate = 0.5              &                     \\ \hline
            \multicolumn{1}{|c|}{Dense}                                       & \multicolumn{1}{c|}{53}                                           & \multicolumn{1}{c|}{53}                                          & \multicolumn{1}{c|}{}                                             & \multicolumn{1}{c|}{}                                        &    Softmax                 &  6,837               \\ \hline
            \multicolumn{1}{l}{}                                              & \multicolumn{1}{l}{}                                              & \multicolumn{1}{l}{}                                             & \multicolumn{1}{l}{}                                              & \multicolumn{1}{l}{}                                         & \cellcolor[HTML]{BBDAFF}\textbf{Total} & \cellcolor[HTML]{BBDAFF}776,341 \\ \cline{6-7} 
            \end{tabular}
            \label{tab:network_generic}
        \end{table*}

        Table \ref{tab:network_generic} shows an alternative network architecture we implemented to demonstrate performance against a more generic architecture that does not have the Temporal-to-Spatial feature enforcement. It also serves to show how early temporal enforcement helps guide network learning as although it is possible for the network to learn temporal features in this design adding temporal enforcement still improves performance. This architecture was also configured to have a similar number of total parameters to our other implementations to help maintain comparability, although on database~2 there is $\sim 20\%$ increase in total parameters. We call this network the Baseline CNN.
        
        Table \ref{tab:network_generic_db2} shows another example of how it is possible to manipulate a network to cover the same time period in the first convolution on database~2 as on database~1 by multiplying the temporal dimension of the filter size by the difference in sampling period (20x) and compensating for this size increase by increasing the stride in the temporal dimension by the same factor. As in our TtS design this allows us to limit the extra parameters added while maintaining performance.
        
        The increase in size along spatial dimensions in the later convolutional filters is necessary to account for increased number of channels in database~2.

         \begin{table*}
            \centering
            \scriptsize
            \caption{Baseline CNN architecture for database~2. Channels last format.}
            \begin{tabular}{cccccc|c|}
            \hline
            \rowcolor[HTML]{6665CD} 
            \multicolumn{1}{|c|}{\cellcolor[HTML]{6665CD}\textbf{Layer Type}} & \multicolumn{1}{c|}{\cellcolor[HTML]{6665CD}\textbf{Output Size}} & \multicolumn{1}{c|}{\cellcolor[HTML]{6665CD}\textbf{\# Filters}} & \multicolumn{1}{c|}{\cellcolor[HTML]{6665CD}\textbf{Filter Size}} & \multicolumn{1}{c|}{\cellcolor[HTML]{6665CD}\textbf{Stride}} &  \textbf{Activation}         &  \textbf{\# Params}          \\ \hline
            \multicolumn{1}{|c|}{Input EMG}                                   & \multicolumn{1}{c|}{300x12x1}                                      & \multicolumn{1}{c|}{}                                            & \multicolumn{1}{c|}{}                                             & \multicolumn{1}{c|}{}                                        &                               &                                           \\ \hline
            \rowcolor[HTML]{BBDAFF} 
            \multicolumn{1}{|c|}{\cellcolor[HTML]{BBDAFF}Gaussian Noise}      & \multicolumn{1}{c|}{\cellcolor[HTML]{BBDAFF}300x12x1}              & \multicolumn{1}{c|}{\cellcolor[HTML]{BBDAFF}}                    & \multicolumn{1}{c|}{\cellcolor[HTML]{BBDAFF}}                     & \multicolumn{1}{c|}{\cellcolor[HTML]{BBDAFF}}                &  $\alpha = 0.001$         &                                   \\ \hline
            \multicolumn{1}{|c|}{Conv}                                        & \multicolumn{1}{c|}{15x12x128}                                    & \multicolumn{1}{c|}{128}                                         & \multicolumn{1}{c|}{60x3}                                          & \multicolumn{1}{c|}{20x1}                                     &  LReLU  &      23,168                         \\ \hline
            \rowcolor[HTML]{BBDAFF} 
            \multicolumn{1}{|c|}{\cellcolor[HTML]{BBDAFF}Conv}                & \multicolumn{1}{c|}{\cellcolor[HTML]{BBDAFF}15x12x64}             & \multicolumn{1}{c|}{\cellcolor[HTML]{BBDAFF}64}                  & \multicolumn{1}{c|}{\cellcolor[HTML]{BBDAFF}5x3}                  & \multicolumn{1}{c|}{\cellcolor[HTML]{BBDAFF}1x1}             & LReLU    &     122,944                         \\ \hline
            \multicolumn{1}{|c|}{Conv}                                        & \multicolumn{1}{c|}{15x12x32}                                     & \multicolumn{1}{c|}{32}                                          & \multicolumn{1}{c|}{5x3}                                          & \multicolumn{1}{c|}{1x1}                                     &   LReLU    &       30,752                          \\ \hline
            \rowcolor[HTML]{BBDAFF} 
            \multicolumn{1}{|c|}{\cellcolor[HTML]{BBDAFF}Dropout}             & \multicolumn{1}{c|}{\cellcolor[HTML]{BBDAFF}4,800}                & \multicolumn{1}{c|}{\cellcolor[HTML]{BBDAFF}}                    & \multicolumn{1}{c|}{\cellcolor[HTML]{BBDAFF}}                     & \multicolumn{1}{c|}{\cellcolor[HTML]{BBDAFF}}                &   rate = 0.5              &                               \\ \hline
            \multicolumn{1}{|c|}{Dense}                                       & \multicolumn{1}{c|}{128}                                          & \multicolumn{1}{c|}{128}                                         & \multicolumn{1}{c|}{}                                             & \multicolumn{1}{c|}{}                                        &    LReLU   &        737,408                      \\ \hline
            \rowcolor[HTML]{BBDAFF} 
            \multicolumn{1}{|c|}{\cellcolor[HTML]{BBDAFF}Dropout}             & \multicolumn{1}{c|}{\cellcolor[HTML]{BBDAFF}128}                  & \multicolumn{1}{c|}{\cellcolor[HTML]{BBDAFF}}                    & \multicolumn{1}{c|}{\cellcolor[HTML]{BBDAFF}}                     & \multicolumn{1}{c|}{\cellcolor[HTML]{BBDAFF}}                &     rate = 0.5             &                                 \\ \hline
            \multicolumn{1}{|c|}{Dense}                                       & \multicolumn{1}{c|}{53}                                           & \multicolumn{1}{c|}{53}                                          & \multicolumn{1}{c|}{}                                             & \multicolumn{1}{c|}{}                                        &   Softmax                &         5,289                           \\ \hline
            \multicolumn{1}{l}{}                                              & \multicolumn{1}{l}{}                                              & \multicolumn{1}{l}{}                                             & \multicolumn{1}{l}{}                                              & \multicolumn{1}{l}{}                                        & \cellcolor[HTML]{BBDAFF}\textbf{Total} & \cellcolor[HTML]{BBDAFF}919,561 \\ \cline{6-7} 
            \end{tabular}
            \label{tab:network_generic_db2}
        \end{table*}

    \subsection{Feature Based Classification (SVM)} 
        We implemented a Support Vector Machine (SVM) as a baseline for comparison to feature based classification on our robust methodology. The SVM used a Radial Basis Function Kernel (RBF) and three different features: the marginal Discrete Wavelet Transform (mDWT) \cite{Lucas2008}, Mean Absolute Value (MAV) \cite{Hudgins} and Waveform Length (WL) \cite{Hudgins}. 
        
        The mDWT for a channel is described as
        \begin{equation} 
            \mathbf{x}_{mdwt} = \sum_{\tau=0}^{N/2^s - 1} \left|\sum_{n=1}^{N} x_n\psi_{l,\tau}(t)\right|
        \end{equation}
        for \( s = 1...S \), where S is the maximum level of decomposition (3 levels were used), $\psi$ is the mother wavelet (sym4 here based on \cite{Atzori2014d}), $l$ is a translation and $\tau$ is a dilation, $x_n$ is the signal value at sample $n$ and $N$ is the length of the signal.
        
        The MAV is described as
        \begin{equation} 
            \mathbf{x}_{mav} = \frac{1}{N} \sum_{n=1}^{N} \left|x_n\right|
        \end{equation}
        
        The WL is described as
        \begin{equation} 
            \mathbf{x}_{wl} = \sum_{n=1}^{N - 1} \left|x_{n+1} - x_{n}\right|
        \end{equation}
        
        SVM feature data was independently normalised after extraction using training set data, class weighting was used based on frequency in the training set (see Equation \ref{eq:class_weight}) and one-vs-all multi-classing was used.

    \subsection{Data Preprocessing} \label{sec:data}
        The open source NinaPro databases 1 and 2 were utilised here \cite{Atzori2014a, Gijsberts2014a, NinaProProject2015}. 
        
        Database~1 contains labelled data from 27 human subjects, performing 10 repetitions of 52 hand movements. The subjects rested between each repetition and so \textit{rest} is treated as an additional movement, leading to 53 movements in total (where \textit{rest} vastly outnumbers all other movements, leading to an imbalanced classification problem). The corresponding sEMG signals were recorded with a 10 channel Otto Bock system, sampled at 100 Hz. This low sampling frequency is due to the root mean square filtering of the Otto Bock electrodes,  which shifts the frequency spectrum to 0-5 Hz \cite{Atzori2014d} from the relevant range for sEMG of 20-500 Hz \cite{DeLuca1997a}. An sEMG signal window of length 150 ms, with 10 ms increment, was used as input to the classifiers. At the sample rate of 100 Hz, this corresponded to windows of 15 samples, with an increment of 1 sample.
        
        Database~2 contains labelled data from 40 human subjects, performing 6 repetitions of 40 hand movements, with \textit{rest} treated as an additional movement, so 41 movements in total. The corresponding sEMG signals were recorded with a 12 channel Delsys system, sampled at 2000 Hz. Database~2 also includes 9 finger-based force pattern exercises that we do not consider here. An sEMG signal window of length 150 ms, with 10 ms increment, was used as input to the classifiers. At the sample rate of 2000 Hz, this corresponded to windows of 300 samples, with an increment of 20 samples.
    
        The data set for an individual subject from either database comprises sEMG input data and corresponding hand movement class labels, 
        \begin{equation} 
            D = \Bigl \lbrace \left(X^{(1)}, y^{(1)} \right), \ldots,  \left(X^{(N)}, y^{(N)}\right) \Bigr \rbrace
        \end{equation}
        where $D$ is the data set for a given subject, the matrix $X^{(j)} \in \mathbb{R}^{n_s \times n_c}$ is a window of sEMG data, with number of samples $n_s$ and number of channels $n_c$, $N$ is the number of data pairs, and $y^{(j)}$ is the corresponding movement label/class,
        \begin{equation} 
             y^{(j)} \in  \mathbb{M} = \lbrace 0,\ldots, M - 1 \rbrace  \text{ for } j=1,\ldots,N
        \end{equation}
        where $M$ is the total number of movements. 

        We used a sliding window method to segment data in both databases with a window length of 150ms and window increment of 10ms. The window length of 150ms was chosen to help ensure classification can occur within an acceptable latency \cite{Farrell2007} and to allow some comparison with the work of Geng \textit{et al} \cite{Geng2016, Du2017} and Atzori \textit{et al} \cite{Atzori2016}. 
        
\begin{figure}
            \centering
            \includegraphics[width=\columnwidth]{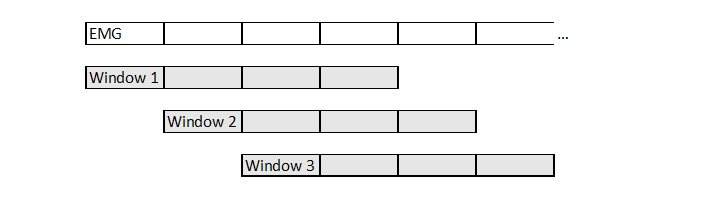}
            \caption{Illustration of sliding window method for an example window length of 4 samples and increment of 1 showing overlap in information between nearby windows.}
            \label{fig:sliding_window}
        \end{figure}

        The sliding window method is illustrated in Fig \ref{fig:sliding_window}. An important issue with this method is that overlapping windows share information which makes it necessary to use an alternative to random selection when dividing into training and testing sets, see Section \ref{sec:validation}.
        
        In the original data all instances of the \textit{rest} class are labelled as being part of an additional repetition 0 however due to the need to split via repetition number (and to keep inline with previous work \cite{Atzori2014d, Atzori2016}) we labelled half the \textit{rest} data before and after each non-\textit{rest} movement as belonging to the repetition of that movement up to a maximum of 10 seconds of \textit{rest} data on either side. To avoid information sharing between windows in different repetitions a gap the size of the window length is also enforced between successive repetitions.
        
        The data is first compiled into a single stream by concatenating the data from each exercise then the windowing starts at the first sample and is slid across the entire data stream (in the increments previously described) with labels for repetition and movement class taken from the most recent sample. The relabelling of \textit{rest} repetitions prevents the discontinuity between exercises being used in training/testing since the data is split into sets based on repetition number and the remaining windows labelled with repetition 0 are ignored.
        
        All data was independently normalised to zero-mean and standard deviation one using training set data for each validation fold.
    
    \subsection{Per Sample Accuracy versus Per Class Accuracy} \label{sec:validation}
        We use two main performance metrics in this paper. The first is a per sample metric that weights samples equally when taking the average performance, termed the micro-average accuracy \cite{Sokolova2009}, which is commonly used elsewhere but is sensitive to class imbalance and so not recommended here. The micro-average is reported only because it has been used elsewhere. The second metric is a per class metric that weights classes equally when taking the average performance, termed the macro-average accuracy \cite{Sokolova2009}, which is not sensitive to class imbalance. Area Under Curve metrics are not used since they are not well defined for the multi-class case \cite{Sokolova2009}.
        
        The micro-average accuracy is defined as
        \begin{equation}
            {Acc}_{micro} =  \frac{\sum^{M}_{i=1}{ \text{TP}_{i}} }{\sum^{ M}_{i=1}{(\text{TP}_{i} + \text{FN}_{i})}}
        \end{equation}
        where TP$_{i}$ (True Positives) is the number of correct classifications for movement $i$, FN$_{i}$ (False Negatives) is the number of instances of movement $i$ that are predicted to be a different movement and $M$ is the number of movement classes. 
        
        Macro-average accuracy is then defined as
        \begin{equation} 
            {Acc}_{macro} = \frac{1}{M} \sum^{M}_{i=1} \left( \frac{\text{TP}_{i}}{\text{TP}_{i} + \text{FN}_{i}} \right)
        \end{equation}
        where the average is taken over the $M$ classes equally weighted.

    \subsection{Cross-Validation}
        The major two studies on these databases we compare against \cite{Atzori2016, Geng2016, Du2017} use a single training-test split in their evaluations specifically repetitions $[2, 5, 7]$ were used for testing and repetitions $[1, 3, 4, 6, 8, 9, 10]$ training on database~1 and repetitions $[2, 5]$ for testing with $[1, 3, 4, 6]$ for training on database~2. Using only a single split in this way negatively impacts the utility of results as a single split can bias results making them unrepresentative of true expected performance. Therefore we use a stratified cross-validation procedure across multiple splits to achieve a more representative result.
        
        
        \begin{table}[!t]
            \renewcommand{\arraystretch}{1.3}
            
            \centering
             \caption{Training and Testing Sets for Multi Split Cross Validation. Each number refers to a movement repetition (for 10 total repetitions in database 1 and 6 repetition in database 2).}
            \begin{tabular}{|c|c|c||c|c|}
            \hline
             &
            \multicolumn{2}{c||}{\textbf{Database 1}} & \multicolumn{2}{c|}{\textbf{Database 2}} \\
            \hline
            \textbf{Split} & \textbf{Training} & \textbf{Testing} & \textbf{Training} & \textbf{Testing}\\
            \hline \hline
            1 & [1, 3, 4, 6, 8, 9, 10]   & [2, 5, 7]  & [1, 3, 4, 6] & [2, 5] \\
            2 & {[1, 2, 3, 5, 7, 9, 10]} & [4, 6, 8]  & [1, 4, 5, 6] & [2, 3] \\
            3 & {[1, 2, 4, 6, 7, 8, 10]} & [3, 5, 9]  & [1, 2, 3, 5] & [4, 6] \\
            4 & {[1, 2, 5, 6, 8, 9, 10]} & [3, 4, 7]  & [1, 2, 4, 6] & [3, 5] \\
            5 & {[2, 3, 4, 6, 8, 9, 10]} & [1, 5, 7]  & [2, 3, 4, 5] & [1, 6] \\
            6 & {[1, 2, 3, 4, 5, 7, 9]}  & [6, 8, 10] & [2, 3, 5, 6] & [1, 4] \\
            7 & {[3, 5, 6, 7, 8, 9, 10]} & [1, 2, 4]  & &\\
            8 & {[1, 2, 4, 5, 7, 8, 9]}  & [3, 6, 10] & &\\
            9 & {[3, 4, 5, 6, 7, 8, 10]} & [1, 2, 9]  & &\\
            10 & {[1, 2, 3, 4, 5, 6, 7]}  & [8, 9, 10] & &\\
            \hline 
            \end{tabular}
            \label{tab:splits}
        \end{table}
            
        Table \ref{tab:splits} shows the splits used here on the two databases: these splits ensure $\sim70\%$ of the data is used for training and $\sim30\%$ for testing (validation) on any single split, for both databases. Selection was also stratified to ensure each repetition was equally represented in the training and test sets to prevent bias due to differences in repetitions. 
        
        
        Regarding cross-validation using repetition splits; due to the usage of a sliding window with overlap, the random selection of windows as typically used in k-fold cross validation would be an inappropriate way to divide into training and test sets. This is because adjacent windows share the majority of their data (see Fig \ref{fig:sliding_window}), thus violating the assumption of independence between training and testing sets. Therefore splitting by repetition number instead ensures proper separation.
        
        Finally we use Forman \textit{et al}'s method to report the final accuracies across validation folds \cite{Forman2010}. This method helps eliminate bias caused by differences in validation folds and is described for the micro-average accuracy as:
        \begin{equation} 
            {Acc}_{micro}^* = \frac{  \sum_{i=1}^M \sum_{j=1}^K \text{TP}_{i,j}}{ \sum^{ M}_{i=1} \sum_{j=1}^K{(\text{TP}_{i, j} + \text{FN}_{i, j})}} 
        \end{equation}
        where $K$ is the number of cross-validation folds, ${TP}_{i,j}$ is the number of correct predictions of movement $i$ for validation fold $j$ and ${FN}_{i, j}$ is the number of instances of movement $i$ that are predicted to be a different movement for validation fold $j$.
        
        For macro-average accuracy the equation becomes:
        \begin{equation} 
            {Acc}_{macro}^* = \frac{1}{M} \sum_{i=1}^M \left( \frac{  \sum_{j=1}^K \text{TP}_{i,j}}{ \sum_{j=1}^K \text{TP}_{i,j} + \sum_{j=1}^K \text{FN}_{i,j}} \right)
        \end{equation}
        
        The final performance for a given classifier, under both metrics, is calculated as the inter subject mean and standard deviation of the current metrics in order to capture the variability between different subjects.
    
    \subsection{Reproduction of Previous Studies} 
        In order to provide a benchmark comparison to the literature directly, we re-implement the networks presented by Atzori \textit{et al} \cite{Atzori2016} and Geng \textit{et al} \cite{Geng2016, Du2017}. 
        
        When re-implementing Geng \textit{et al}'s network we utilised their available code \cite{Geng, Du2017} and retested using their own methodology to ensure correctness. We found a $<1\%$ difference in performance between our implementations which may be caused by differences in supporting software/hardware and random number seeding which was not set explicitly in their implementation. When running this network with our validation procedures we used majority voting over each of the windows to determine the final predicted output as used in their study. 
        
        The key differences in methodology between our work and Geng \textit{et al}'s \cite{Geng2016, Du2017} are that Geng \textit{et al} omit the \textit{rest} class from consideration, compare performance based on micro-average accuracy, validate on a single split and only classify 8 finger force exercises on database~2 rather than the 41 hand movements that we classify.
        
        For the Atzori \textit{et al} network \cite{Atzori2016} we re-implement their network based on their description. In their paper several parameters were unspecified, including convolution stride and padding, therefore we tested a small pool of potential parameters and used the best performing among them. Specifically we assumed padding was used to maintain shape (as in our networks), both pooling layers used stride equal to their size (3x3), Block 3's convolutional layer used a stride of 5x5 and Block 4 used a stride of 9x1 on database~2. These parameters allowed all the specified parameters to remain the same while accounting for shape changes necessary for operation.
        
        Our validation methods are similar to Atzori \textit{et al} with the major differences being our use of multiple splits in validation (versus their single split) and choice of performance metrics. In Atzori \textit{et al} \cite{Atzori2016} an earlier paper was referenced  as the methodological base \cite{Atzori2014d}: this earlier paper uses the micro-average accuracy as its metric without data balancing however it is stated in \cite{Atzori2016} that the data was balanced by repetition number. This removes the large skew towards rest however does not take account of the difference between other classes which on this data causes some classes to be weighted as up to 2x more important than other classes (section \ref{sec:validation}) leading to bias in the result.
        
        The other main difference between our work and both these studies is that we do not perform the additional preprocessing step of zero-phase low-pass filtering. We chose to omit this as it is not possible perform true zero-phase filtering in online contexts which is a primary use-case for sEMG classification methods making it an inappropriate method to use when benchmarking.
        
    \subsection{Statistical Comparison of Classifiers}  \label{sec:stats}
        Accepted best practice for statistical testing of multiple classifiers over multiple data sets is well defined by Dem\u{s}ar \cite{Dem2006}. Here each subject is a different data set since each classifier is trained independently on each subject. In order to demonstrate that our TtS design significantly improves over other classifiers we take the approach recommended by Dem\u{s}ar \cite{Dem2006}, of using the Friedman test \cite{Friedman1937}, with Iman and Davenport's improved statistic \cite{Iman1980}, to establish that the pool of classifiers under investigation show different performances and then the post-hoc Holm Procedure \cite{Holm1979}, to confirm that the TtS network improves upon each other classifier. We performed the Holm Procedure separately on each Ninapro database and use $2\%$ as our significance level.
        
        The Holm Procedure first calculates the $p$ value for each pair of interest (TtS classifier vs another classifier) based upon the average performance rank calculated by the Friedman test. These $p$ values are then sorted in ascending order (most significant value first). Then the procedure operates in a step down fashion, at the first step the significance level $\alpha$ is reduced by $\alpha / (k-1)$ where $k$ is the number of classifiers being investigated (here $k=5$) to account for the number of comparisons that may occur. If $p < \alpha$ then the null hypothesis is rejected and we are allowed to compare to the next most significant $p$ value with $\alpha / (k-2)$ and this process is repeated up to $k-1$ times. If any null hypothesis cannot be rejected then the process is stopped and all remaining null hypotheses are kept as well.
        
        The full experimental process, from data to statistical analysis, is illustrated in Fig \ref{fig:expt_design}.
        
        \begin{figure}
            \centering
            \caption{Experimental process flowchart.}
            \label{fig:expt_design}
        \end{figure}

\section{Results}      

Our proposed Temporal-to-Spatial (TtS) CNN was compared to our re-implementations of two published CNNs by Geng et al. \cite{Geng2016} and Atzori et al. \cite{Atzori2016}, as well as a feature-based classifier, an SVM with RBF kernel, and a baseline CNN of our own design with no TtS structure. The results are summarised in Table \ref{tab:res_db1}, which show that the TtS network outperformed all other classifiers on both Ninapro databases 1 and 2 in terms of macro-average accuracy (which is a measure of performance that equally weights all classes). The results are given as the inter-subject mean and standard deviation to show expected performance on a new subject. Results are ordered by the mean macro-average accuracy since it is more representative of expected performance on this data (see Section \ref{sec:metric_good}).

Following Dem\u{s}ar \cite{Dem2006} we used the Friedman test \cite{Friedman1937} (see Section \ref{sec:stats}) to confirm rejection of the null hypothesis that all classifiers performed the same ($p<1$x$10^{-31}$ for both databases). We then used the recommended post-hoc Holm Procedure \cite{Holm1979}, which confirmed that our TtS network performed significantly better than each other classifier on both databases at the $2\%$ significance level.

\begin{figure*}
    \centering
    \includegraphics[scale=0.8]{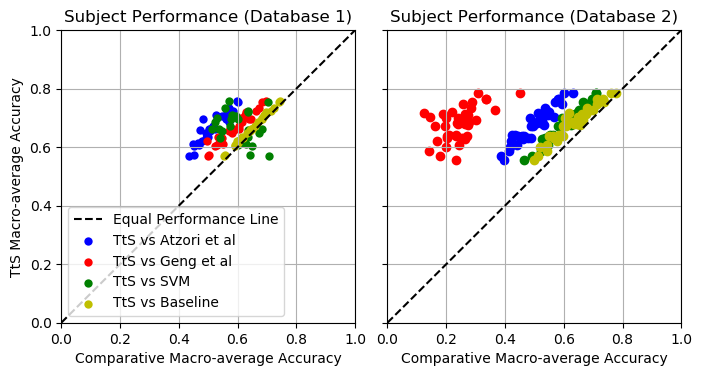}
    \caption{Per subject comparison of performance between the TtS network and its competitors, where each dot represents classifier performance on a single human subject. The results demonstrate that for almost any given subject on either database the TtS network outperforms all the other tested classifiers. The only exception is the SVM on database~1 which outperforms the TtS on 8 of the 27 subjects.}
    \label{fig:perf_compare}
\end{figure*}

Fig \ref{fig:perf_compare} supplements the table demonstrating the relative performance of the TtS network against its competitors for each subject to highlight its consistent performance enhancement.
    
    \begin{table}
        \centering
        \caption{Summary of classifier accuracies using our re-implementations on our methodology. *Repetition 1 removed from test set without retraining.}
        \begin{tabular}{|l|c|c|c|c|}
        \hline
        \multicolumn{5}{|c|}{\textbf{Database 1}}                                                                                      \\ \hline
        \multirow{2}{*}{\textbf{Classifier}}  & \multicolumn{2}{c|}{\textbf{Per Class Acc. (\%)}}  & \multicolumn{2}{c|}{\textbf{Per Sample Acc. (\%)}}    \\ 
                                                  & \textbf{Mean}           & \textbf{Std.} & \textbf{Mean}           & \textbf{Std.} \\ \hline
        Atzori \textit{et al}\cite{Atzori2016}     & 51.4                  & 4.8         & 71.0                  & 4.6         \\ \hline
        Geng \textit{et al} \cite{Geng2016, Du2017}& 58.9                  & 5.7         & 79.9                  & 3.6         \\ \hline
        SVM                                        & 60.4                  & 5.4         & 77.8                  & 4.1         \\ \hline
        Baseline CNN                               & 65.0                  & 5.1         & 77.1                  & 4.7         \\ \hline
        TtS                                        & 66.6                  & 5.1         & 77.5                  & 4.5         \\ \hline
        TtS*                                       & 69.3                  & 5.4         & 78.0                  & 4.6         \\ \hline
        \end{tabular}
        \vspace{1.5ex}
        \label{tab:res_db1}
        \centering
        \begin{tabular}{|l|c|c|c|c|}
        \hline
        \multicolumn{5}{|c|}{\textbf{Database 2}}                                                                                      \\ \hline
        \multirow{2}{*}{\textbf{Classifier}}  & \multicolumn{2}{c|}{\textbf{Per Class Acc. (\%)}}  & \multicolumn{2}{c|}{\textbf{Per Sample Acc. (\%)}}    \\ 
                                                  & \textbf{Mean}           & \textbf{Std.} & \textbf{Mean}           & \textbf{Std.} \\ \hline
        Geng \textit{et al} \cite{Geng2016, Du2017}        & 24.5                  & 6.3         & 58.2                  & 8.3         \\ \hline
        Atzori \textit{et al}\cite{Atzori2016}     & 50.3                  & 5.9         & 61.3                  & 8.1         \\ \hline
        Baseline CNN                               & 57.0                  & 6.2         & 64.8                  & 8.3         \\ \hline
        SVM                                        & 60.5                  & 6.3         & 71.2                  & 6.8         \\ \hline
        TtS                                        & 67.8                  & 5.7         & 69.5                  & 7.8         \\ \hline
        TtS*                                       & 70.6                  & 6.1         & 70.9                  & 7.6      \\ \hline
        \end{tabular}
        \label{tab:res_db2}
    \end{table}
    
Micro-average is also reported in Table \ref{tab:res_db1}, despite being an imbalanced metric that is unrepresentative of performance across classes, to allow comparison with previous work which uses the metric. Our results are consistent with previous findings with the micro-average accuracy being similar to the results of Geng \textit{et al} \cite{Geng2016, Du2017} and Atzori \textit{et al} \cite{Atzori2016} for their respective networks. The differences found here are due to our more robust validation procedure which uses stratification and multiple splits to account for bias from different splits. We also note that our SVM implementation performs better than Atzori \textit{et al}'s network under micro-average accuracy which is consistent with their results \cite{Atzori2016}.
    
We also tested the networks for computational feasibility. On a high end computer equipped with an NVIDIA GTX1080Ti GPU, AMD Ryzen 1700 CPU and DDR4-2933MHz RAM all networks took, on average, 1-4 ms to perform a forward pass on a single data point.

\section{Discussion}   \label{sec:discussion}
    \subsection{Representative Performance Evaluation}  \label{sec:metric_good}

One of the most important factors in comparing performance is the metric used. The performance metric in many related studies is either not directly specified, or is the micro-average accuracy \cite{Atzori2016, Shin2014, Geng2016, Du2017, Ju2013, Khushaba2009}. However it is known that the micro-average accuracy is sensitive to class imbalance which means its usage on imbalanced data will lead to unrepresentative results \cite{Sokolova2009} particular when classifying EMG data \cite{Ortiz-Catalan2015}. In the databases under investigation the data is highly imbalanced with there being many more examples of \textit{rest} than any other class and some non-\textit{rest} classes having over twice as many examples as other non-\textit{rest} classes. Therefore using the micro-average accuracy on this data leads to a skewed, unrepresentative result with some classes being weighted much more than others based, effectively on how long they took to perform since longer movements lead to more data on the class.
        
\begin{figure}
\centering
\includegraphics[width=\columnwidth]{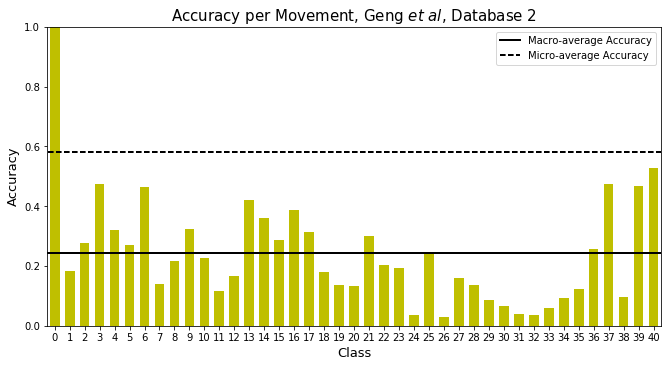}
\caption{Accuracy of individual movement classes compared to the overall micro-average and macro-average accuracy for Geng \textit{et al's} classifier \cite{Geng2016} on database~2 evaluated using our methodology. The lines indicate the large difference between reported performance under the micro-average and macro-average metrics.}
\label{fig:geng_performance}
\end{figure}

The issue is most egregious when contrasting \textit{rest} and non-\textit{rest} classes as for most subjects there is more data on the \textit{rest} classes than all other classes combined. This makes the micro-average accuracy effectively a measure of performance on the \textit{rest} class since over $50\%$ of the variance of the micro-average accuracy is dictated by performance on \textit{rest} class. Fig \ref{fig:geng_performance} demonstrates how this leads to an unrepresentative metric on this data in particular; the micro-average accuracy reports a performance higher than the performance of any class other than \textit{rest}.
        
The macro-average accuracy alleviates the issue of skewed results due to the aforementioned class imbalance by weighting performance on each class equally rather than based on the number of examples in the test set (which is unlikely to be representative of any particular application). Therefore the macro-average accuracy is recommended as the default metric for comparison of performance for the classification of hand movements from sEMG since it is robust to the issue of class imbalance and better represents the overall performance of a classifier particularly as the number of classes increases.

Note as a final point that confusion matrices are often used to analyse classification results \cite{Yang2020}. However, in this case the large number of classes in database 1 (53 classes) and database 2 (41 classes) makes visualisation of the confusion matrices not particularly effective and so has been avoided here.

        
\subsection{Comparison to Alternative Classification Methods}

To compare to alternative classification methods we re-implemented deep CNNs from Geng \textit{et al.} \cite{Geng2016} and Atzori \textit{et al.} \cite{Atzori2016}, and an SVM feature-based classifier similar to that used in \cite{Atzori2014d}. To summarise, we found that our TtS network outperformed all of these methods in terms of macro-average accuracy ($66.6\%$ database~1, $67.8\%$ database~2, Table VI) and that this performance improvement was also confirmed by statistical hypothesis testing, at the $2\%$ significance level (see Results).

        
Compared to our re-implementation of the Geng \textit{et al} network \cite{Geng2016}, the TtS network improved on database~1 windowed classification performance by $\sim7.7\%$ macro-average accuracy, while maintaining a slightly lower standard deviation between subjects. On database~2 the performance of their network decreased significantly to $24.5\%$ ($15.4\%$ in instantaneous mode) which is likely because the network does not account for the $20$x higher sampling frequency leading to $20$x denser data. A smaller, although still significant, drop off also occurred in terms of micro-average accuracy.
        
        
The paper by Geng \textit{et al} \cite{Geng2016} also reported performance over each movement trial, that is: majority voting over all segments known to be from the same movement. We found that, in this setup, their network achieved $87.0\%$ macro-average accuracy on database~1 and $20.1\%$ on database~2, whilst the TtS network achieved $92.9\%$ on database~1 and $95.0\%$ on database~2. We suggest this trial based performance measure, however, is impractical since it is likely to induce significant latency as movements can last up to $5000$ ms which is likely to push the classification latency above the $\sim200$ ms maximum acceptable control delay latency \cite{Farrell2007}. Further, in a practical context this would require a strong prior on when a subject begins and end a movement of interest which is not generally available. 
        
Compared to our re-implementation of Atzori \textit{et al}'s \cite{Atzori2016} CNN, the TtS network improved on database~1  classification performance by $\sim15.2\%$ macro-average accuracy and $\sim 17.5\%$ on database~2. 

The feature-based classifier, an SVM with RBF kernel, achieved a macro-average accuracy on database 1 of $60.4\%$ and $60.5\%$ on database~2. On database~2 it was the second top performer behind our TtS network. The consistency of the SVM is likely due to the feature extraction methods effectively reducing the data in both databases to a similar feature space, combined with leveraging of the SVM's ability to construct an optimal hyper-plane to divide the space.

        

 
We suggest the improvement of the TtS design over the other CNN methods is due to the reason that the EMG data in each channel are mediated by one or a small subset of muscles, whose temporal activation patterns are associated with the signature features of a gesture. The temporal layer captures these signature features for each channel separately. Once these signature features for each channel are captured they are spatially mixed to recognise a specific gesture. This is the key design difference that allowed us to produce the significant performance improvement demonstrated in this paper.
        
\subsection{Computational complexity}

Deep CNNs can be computationally intensive to implement. The computational complexity of a convolutional layer is $O(n_i n_o r_l c_l R_l C_l)$, where $n_i$ is the number of input feature maps, $n_o$ is the number of output feature maps, $r_l\times c_l$ is the feature map size, and $R_l \times C_l$ is the size of the convolution filter \cite{Cong2014}. In practice, the computation time is limited by the number of cores in the GPU used for implementation (as well as other factors such as memory bandwidth).

We found here that on a high performance workstation GPU, an NVIDIA Geforce 1080Ti, the computation time was on the order of 1-4 ms to process a single forward pass through the TtS network, which is likely to be sufficient for real-time implementation. Of more relevance to embedded systems, CNNs of a similar design have been implemented by the authors in \cite{Hartwell2018}, on an NVIDIA Jetson Tx2 (embedded system), with times of around 20 ms for a single forward pass that can be reduced to $\sim 8$ ms using network compression. This suggests that deep CNNs can be implemented currently at usable sample rates in both modern computational settings and embedded systems.

        
\subsection{Effect of First Repetition}

\begin{figure}
            \centering
            \includegraphics[width=0.9\columnwidth]{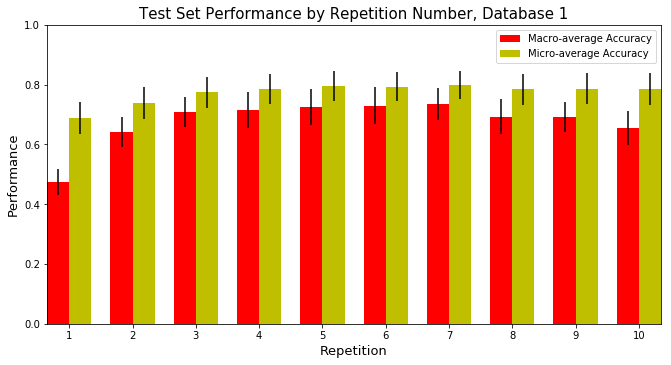}
            \includegraphics[width=0.9\columnwidth]{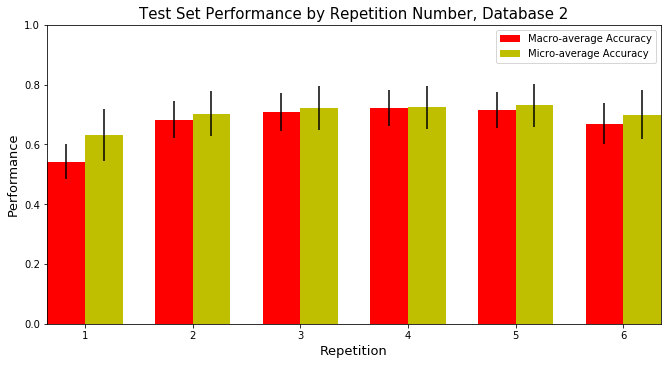}
            \caption{Performance by repetition of the TtS network on the two databases demonstrating a distinct performance reduction in the first repetition for both databases.}
            \label{fig:repetition_performance}
        \end{figure}

During analysis we found that, on average, the first repetition of each movement each subject performed had a lower classification rate than the later repetitions of the movement (Fig. \ref{fig:repetition_performance}). The magnitude of this effect is shown in Table IV which shows $\sim3\%$ performance improvement when repetition 1 is not considered in testing, dropping other repetitions leads to a much smaller effect on the reported performance. We suggest that this is likely an artefact of the experimental procedure: the first, or first few times a subject performs a movement or after having performed another movement there is a higher likelihood of error. This is backed up by the observation from Fig \ref{fig:repetition_performance} that the first repetition is noticeably worse than the others.
    
 To a lesser degree the later repetitions on database~1 also show a decline in performance. This is unlikely to be from muscle fatigue (due to the limited number of repetitions \cite{Farrell2008}) but may be caused by preemption of the video stimuli or inattention causing non-optimal replication of the movement. Alternatively this may be explained by user adaptation during periods of consistent training, an effect that would be amplified by classifier-feedback technology if it is used.

\section{Conclusion} \label{Conclusions}
    We have developed a novel deep CNN, with a Temporal-to-Spatial (TtS) architecture, for hand movement classification from surface EMG signals. The TtS architecture constrains early convolutions to only expand along the temporal dimension which enforces learning of temporal features (rather than cross-channel spatial features) as the basis of classification. We also introduce the Temporal Fire Module which is based on the SqueezeNet architecture which improves performance for a minimal cost in number of parameters. Lastly our design includes a simple solution to account for vastly different sampling frequencies or changes in window length in different sEMG data sets which requires minimal architectural changes while maintaining performance. Cross-validation and statistical comparison of classifiers demonstrated that the proposed TtS network outperformed previous CNN designs and a feature-based classifier, an SVM with RBF kernel. These results suggest that the TtS network design is particularly effective for EMG based hand gesture classification.
    
    

    \subsection{Future Work}
    While we demonstrate significant improvements over previous work, our best macro-average accuracy across subjects is $70.6\%$, which indicates that classification of 40+ movements is not necessarily practical. Therefore, a primary avenue for future research is further improvements to either the classification approach or data acquisition methods in order to improve overall performance.
    
    During data acquisition, careful choice of electrode size and position can help improve performance \cite{Young2011}. Sub-selection of movements on a person-specific basis from a larger set of movements like the NinaPro Databases can help improve performance when not all movements need to be classified simultaneously \cite{Hartwell}. Similarly, sensor fusion methods could also be used to improve CNN performance, incorporating other sensor data along with the sEMG such as accelerometers \cite{Zhang2011, Lu2014}, inertial measurement units \cite{Jiang2016, Krasoulis2017} and/or near-infrared spectroscopy \cite{Guo2017}. 
    
    Part of the deployment procedure could incorporate classifier-feedback user training regimes \cite{Fang2017} which may also improve performance without the need for changes in the CNN or data acquisition strategy.
    
    Alternatively, extension of these network architectures to other facets of the muscle-computer interface problem, such as force-related prosthetic control \cite{Yang2019}, could lead to improvements in the performance of interface devices.

\bibliography{library.bib}

\begin{thebibliography}{10}

\bibitem{Mitra2007}
Mitra S, Acharya T.
\newblock {Gesture recognition: A survey}.
\newblock IEEE Transactions on Systems, Man, and Cybernetics, Part C
  (Applications and Reviews). 2007;37(3):311--324.

\bibitem{AsghariOskoei2007}
{Asghari Oskoei} M, Hu H.
\newblock {Myoelectric control systems-A survey}.
\newblock Biomedical Signal Processing and Control. 2007;2(4):275--294.
\newblock doi:{10.1016/j.bspc.2007.07.009}.

\bibitem{Farina2014}
Farina D, Jiang N, Rehbaum H, Holobar A, Graimann B, Dietl H, et~al.
\newblock {The extraction of neural information from the surface EMG for the
  control of upper-limb prostheses: emerging avenues and challenges}.
\newblock IEEE Transactions on Neural Systems and Rehabilitation Engineering.
  2014;22(4):797--809.

\bibitem{Huang2017}
Huang Y, Yang X, Li Y, Zhou D, He K, Liu H.
\newblock {Ultrasound-based sensing models for finger motion classification}.
\newblock IEEE Journal of Biomedical and Health Informatics.
  2017;22(5):1395--1405.

\bibitem{Patsadu2012}
Patsadu O, Nukoolkit C, Watanapa B.
\newblock {Human gesture recognition using Kinect camera}.
\newblock In: 2012 ninth international conference on computer science and
  software engineering (JCSSE). IEEE; 2012. p. 28--32.

\bibitem{Ju2009}
Ju Z, Liu H, Zhu X, Xiong Y.
\newblock {Dynamic grasp recognition using time clustering, {G}aussian mixture
  models and hidden {M}arkov models}.
\newblock Advanced Robotics. 2009;23(10):1359--1371.

\bibitem{Duan2018}
Duan F, Ren X, Yang Y.
\newblock {A gesture recognition system based on time domain features and
  linear discriminant analysis}.
\newblock IEEE Transactions on Cognitive and Developmental Systems. 2018;.

\bibitem{Atzori2014d}
Atzori M, Gijsberts A, Kuzborskij I, Elsig S, {Mittaz Hager} AG, Deriaz O,
  et~al.
\newblock {Characterization of a Benchmark Database for Myoelectric Movement
  Classification}.
\newblock IEEE Transactions on Neural Systems and Rehabilitation Engineering.
  2015;23(1):73--83.
\newblock doi:{10.1109/TNSRE.2014.2328495}.

\bibitem{Castellini2009}
Castellini C, {Van Der Smagt} P.
\newblock {Surface EMG in advanced hand prosthetics}.
\newblock Biological Cybernetics. 2009;100(1):35--47.

\bibitem{Quitadamo2017}
Quitadamo LR, Cavrini F, Sbernini L, Riillo F, Bianchi L, Seri S, et~al.
\newblock {Support vector machines to detect physiological patterns for EEG and
  EMG-based human-computer interaction: a review}.
\newblock Journal of Neural Engineering. 2017;14(1):11001.

\bibitem{Duan2016}
Duan F, Dai L, Chang W, Chen Z, Zhu C, Li W.
\newblock {sEMG-based identification of hand motion commands using wavelet
  neural network combined with discrete wavelet transform}.
\newblock IEEE Transactions on Industrial Electronics. 2016;63(3):1923--1934.

\bibitem{Balbinot2013}
Balbinot A, Favieiro G.
\newblock {A neuro-fuzzy system for characterization of arm movements}.
\newblock Sensors. 2013;13(2):2613--2630.

\bibitem{Khezri2011}
Khezri M, Jahed M.
\newblock {A neuro-fuzzy inference system for sEMG-based identification of hand
  motion commands}.
\newblock IEEE Transactions on Industrial Electronics. 2011;58(5):1952--1960.

\bibitem{Baldacchino2018}
Baldacchino T, Jacobs WR, Anderson SR, Worden K, Rowson J.
\newblock {Simultaneous Force Regression and Movement Classification of Fingers
  via Surface EMG within a Unified Bayesian Framework}.
\newblock Frontiers in Bioengineering and Biotechnology. 2018;6:13.

\bibitem{Shin2014}
Shin S, Tafreshi R, Langari R.
\newblock {A performance comparison of hand motion EMG classification}.
\newblock In: Proceedings of the 2nd Middle East Conference on Biomedical
  Engineering. IEEE; 2014. p. 353--356.

\bibitem{Khushaba2009}
Khushaba RN, Al-Jumaily A, Al-Ani A.
\newblock {Evolutionary fuzzy discriminant analysis feature projection
  technique in myoelectric control}.
\newblock Pattern Recognition Letters. 2009;30(7):699--707.

\bibitem{Khezri2007}
Khezri M, Jahed M.
\newblock {Real-time intelligent pattern recognition algorithm for surface EMG
  signals}.
\newblock Biomedical Engineering Online. 2007;6(1):1.

\bibitem{Huang2005}
Huang Y, Englehart KB, Hudgins B, Chan ADC.
\newblock {A Gaussian mixture model based classification scheme for myoelectric
  control of powered upper limb prostheses}.
\newblock IEEE Transactions on Biomedical Engineering. 2005;52(11):1801--1811.

\bibitem{Lucas2008a}
Lucas MF, Gaufriau A, Pascual S, Doncarli C, Farina D.
\newblock {Multi-channel surface EMG classification using support vector
  machines and signal-based wavelet optimization}.
\newblock Biomedical Signal Processing and Control. 2008;3(2):169--174.
\newblock doi:{10.1016/j.bspc.2007.09.002}.

\bibitem{Ju2013}
Ju Z, Ouyang G, Wilamowska-Korsak M, Liu H.
\newblock {Surface EMG based hand manipulation identification via nonlinear
  feature extraction and classification}.
\newblock IEEE Sensors Journal. 2013;13(9):3302--3311.

\bibitem{Khushaba2007}
Khushaba RN, Al-Jumaily A.
\newblock {Fuzzy wavelet packet based feature extraction method for
  multifunction myoelectric control}.
\newblock International Journal of Biological and Medical Sciences.
  2007;2(3):186--194.

\bibitem{Tenore2009}
Tenore FVG, Ramos A, Fahmy A, Acharya S, Etienne-Cummings R, Thakor NV.
\newblock {Decoding of individuated finger movements using surface
  electromyography}.
\newblock IEEE Transactions on Biomedical Engineering. 2009;56(5):1427--1434.

\bibitem{Zhou2010}
Zhou R, Liu X, Li G.
\newblock {Myoelectric signal feature performance in classifying motion classes
  in transradial amputees}.
\newblock In: Proceedings of the Congress of the International Society of
  Electrophysiology and Kinesiology; 2010. p. 16--19.

\bibitem{Ortiz-Catalan2012}
Ortiz-Catalan MJ, Br{\aa}nemark R, H{\aa}kansson B.
\newblock {Biologically inspired algorithms applied to prosthetic control}.
\newblock In: BioMed 2012; 2012. p. 764.

\bibitem{Canal2010}
Canal MR.
\newblock {Comparison of wavelet and short time Fourier transform methods in
  the analysis of EMG signals}.
\newblock Journal of Medical Systems. 2010;34(1):91--94.

\bibitem{Reaz2006}
Reaz MBI, Hussain MS, Mohd-Yasin F.
\newblock {Techniques of EMG signal analysis: detection, processing,
  classification and applications}.
\newblock Biological Procedures Online. 2006;8(1):11--35.

\bibitem{LeCun2015}
LeCun Y, Bengio Y, Hinton G, Y L, Y B, G H.
\newblock {Deep learning}.
\newblock Nature. 2015;521(7553):436--444.
\newblock doi:{10.1038/nature14539}.

\bibitem{Atzori2016}
Atzori M, Cognolato M, M{\"{u}}ller H.
\newblock {Deep learning with convolutional neural networks applied to
  electromyography data: A resource for the classification of movements for
  prosthetic hands}.
\newblock Frontiers in Neurorobotics. 2016;10(SEP):1--10.
\newblock doi:{10.3389/fnbot.2016.00009}.

\bibitem{Zhai2017}
Zhai X, Jelfs B, Chan RHM, Tin C.
\newblock {Self-recalibrating surface EMG pattern recognition for
  neuroprosthesis control based on convolutional neural network}.
\newblock Frontiers in Neuroscience. 2017;11:379.

\bibitem{Ameri2019}
Ameri A, Akhaee MA, Scheme E, Englehart K.
\newblock {Regression convolutional neural network for improved simultaneous
  EMG control}.
\newblock Journal of Neural Engineering. 2019;16(3):36015.

\bibitem{Geng2016}
Geng W, Du Y, Jin W, Wei W, Hu Y, Li J.
\newblock {Gesture Recognition by Instantaneous Surface EMG Images}.
\newblock Scientific Reports. 2016;6:36571.

\bibitem{ZiaurRehman2018}
Zia-ur Rehman M, Waris A, Gilani S, Jochumsen M, Niazi I, Jamil M, et~al.
\newblock {Multiday EMG-Based Classification of Hand Motions with Deep Learning
  Techniques}.
\newblock Sensors. 2018;18(8):2497.

\bibitem{Phinyomark2013}
Phinyomark A, Quaine F, Charbonnier S, Serviere C, Tarpin-Bernard F, Laurillau
  Y.
\newblock {EMG feature evaluation for improving myoelectric pattern recognition
  robustness}.
\newblock Expert Systems with Applications. 2013;40(12):4832--4840.
\newblock doi:{10.1016/j.eswa.2013.02.023}.

\bibitem{Iandola2016}
Iandola FN, Moskewicz MW, Ashraf K, Han S, Dally WJ, Keutzer K.
\newblock {SqueezeNet: AlexNet-Level Accuracy with 50x Fewer Parameters and
  {\textless}1MB Model Size}.
\newblock arXiv:160207360. 2016;doi:{10.1007/978-3-319-24553-9}.

\bibitem{Atzori}
Atzori M, Gijsberts A, Heynen S, Hager AgM, Deriaz O, {Van Der Smagt} P, et~al.
\newblock {Building the NINAPRO Database : A Resource for the Biorobotics
  Community}.
\newblock Biomedical Robotics and Biomechatronics. 2012; p. 1258 -- 1265.

\bibitem{Atzori2014a}
Atzori M, Gijsberts A, Castellini C, Caputo B, Hager AGM, Elsig S, et~al.
\newblock {Electromyography data for non-invasive naturally-controlled robotic
  hand prostheses.}
\newblock Scientific Data. 2014;doi:{10.1038/sdata.2014.53}.

\bibitem{NinaProProject2015}
{NinaPro Project}. {Ninapro Database Website}; 2015.
\newblock Available from: \url{http://ninapro.hevs.ch/}.

\bibitem{Du2017}
Du Y, Jin W, Wei W, Hu Y, Geng W.
\newblock {Surface EMG-based inter-session gesture recognition enhanced by deep
  domain adaptation}.
\newblock Sensors. 2017;17(3):458.

\bibitem{Lucas2008}
Lucas MF, Gaufriau A, Pascual S, Doncarli C, Farina D.
\newblock {Multi-channel surface EMG classification using support vector
  machines and signal-based wavelet optimization}.
\newblock Biomedical Signal Processing and Control. 2008;3(2):169--174.

\bibitem{Hudgins}
Hudgins B, Parker P, Robert N.
\newblock {A New Strategy for Multifunction Myoelectric Control}.
\newblock IEEE Transactions on Biomedical Engineering. 1993;40(1):82--94.

\bibitem{Sokolova2009}
Sokolova M, Lapalme G.
\newblock {A systematic analysis of performance measures for classification
  tasks}.
\newblock Information Processing and Management. 2009;45(4):427--437.
\newblock doi:{10.1016/j.ipm.2009.03.002}.

\bibitem{Ortiz-Catalan2015}
Ortiz-Catalan M, Rouhani F, Br{\aa}nemark R, H{\aa}kansson B.
\newblock {Offline accuracy: a potentially misleading metric in myoelectric
  pattern recognition for prosthetic control}.
\newblock In: Proceedings of the 37th Annual International Conference of the
  IEEE Engineering in Medicine and Biology Society. IEEE; 2015. p. 1140--1143.

\bibitem{Dem2006}
Dem{\v{s}}ar J.
\newblock {Statistical Comparisons of Classifiers over Multiple Data Sets}.
\newblock Journal of Machine Learning Research. 2006;7:1--30.
\newblock doi:{10.1016/j.jecp.2010.03.005}.

\bibitem{maas2013rectifier}
Maas AL, Hannun AY, Ng AY.
\newblock {Rectifier nonlinearities improve neural network acoustic models}.
\newblock In: Proceedings of the International Conference on Machine Learning.
  vol.~30; 2013.

\bibitem{Xu2015}
Xu B, Wang N, Chen T, Li M.
\newblock {Empirical evaluation of rectified activations in convolutional
  network}.
\newblock arXiv:150500853. 2015;.

\bibitem{Glorot2010}
Glorot X, Bengio Y.
\newblock {Understanding the difficulty of training deep feedforward neural
  networks}.
\newblock Proceedings of the 13th International Conference on Artificial
  Intelligence and Statistics. 2010;9:249--256.
\newblock doi:{10.1.1.207.2059}.

\bibitem{chollet2015keras}
{F  Chollet and others}. {Keras}; 2015.
\newblock Available from: \url{https://github.com/fchollet/keras}.

\bibitem{Abadi2016}
Abadi M, Barham P, Chen J, Chen Z, Davis A, Dean J, et~al.
\newblock {TensorFlow: A System for Large-Scale Machine Learning.}
\newblock In: OSDI. vol.~16; 2016. p. 265--283.

\bibitem{Chen2015}
Chen T, Li M, Li Y, Lin M, Wang N, Wang M, et~al.
\newblock {Mxnet: A flexible and efficient machine learning library for
  heterogeneous distributed systems}.
\newblock arXiv:151201274. 2015;.

\bibitem{Scheme2014}
Scheme E, Englehart K.
\newblock {On the robustness of EMG features for pattern recognition based
  myoelectric control; a multi-dataset comparison}.
\newblock In: Proceedings of the 36th Annual International Conference of the
  IEEE Engineering in Medicine and Biology Society. IEEE; 2014. p. 650--653.

\bibitem{Pizzolato2017}
Pizzolato S, Tagliapietra L, Cognolato M, Reggiani M, M{\"{u}}ller H, Atzori M.
\newblock {Comparison of six electromyography acquisition setups on hand
  movement classification tasks}.
\newblock PLOS ONE. 2017;12(10):e0186132.

\bibitem{Kingma2015}
Kingma DP, Ba JL.
\newblock {Adam: a Method for Stochastic Optimization}.
\newblock International Conference on Learning Representations. 2015; p. 1--15.

\bibitem{Gijsberts2014a}
Gijsberts A, Atzori M, Castellini C, M{\"{u}}ller H, Caputo B.
\newblock {Movement Error Rate for Evaluation of Machine Learning Methods for
  sEMG-Based Hand Movement Classification}.
\newblock IEEE Transactions on Neural Systems and Rehabilitation Engineering.
  2014;22(4):735--744.
\newblock doi:{10.1109/TNSRE.2014.2303394}.

\bibitem{DeLuca1997a}
{De Luca} CJ.
\newblock {The use of surface electromyography in biomechanics}.
\newblock Journal of Applied Biomechanics. 1997;13:135--163.
\newblock doi:{citeulike-article-id:2515246}.

\bibitem{Farrell2007}
Farrell TR, Weir RF.
\newblock {The optimal controller delay for myoelectric prostheses}.
\newblock IEEE Transactions on Neural Systems and Rehabilitation Engineering.
  2007;15(1):111--118.

\bibitem{Forman2010}
Forman G, Scholz M.
\newblock {Apples-to-apples in cross-validation studies}.
\newblock ACM SIGKDD Explorations Newsletter. 2010;12(1):49.
\newblock doi:{10.1145/1882471.1882479}.

\bibitem{Geng}
Geng W, Du Y, Jin W, Wei W, Hu Y, Li J. {Gesture Recognition by Instantaneous
  Surface EMG Images [code]}; 2016.
\newblock Available from: \url{http://zju-capg.org/myo/}.

\bibitem{Friedman1937}
Friedman M.
\newblock {The use of ranks to avoid the assumption of normality implicit in
  the analysis of variance}.
\newblock Journal of the American Statistical Association.
  1937;32(200):675--701.

\bibitem{Iman1980}
Iman RL, Davenport JM.
\newblock {Approximations of the critical region of the fbietkan statistic}.
\newblock Communications in Statistics-Theory and Methods. 1980;9(6):571--595.

\bibitem{Holm1979}
Holm S.
\newblock {A simple sequentially rejective multiple test procedure}.
\newblock Scandinavian Journal of Statistics. 1979; p. 65--70.

\bibitem{Yang2020}
Yang Y, Duan F, Ren J, Xue J, Lv Y, Zhu C, et~al.
\newblock {Performance Comparison of Gestures Recognition System Based on
  Different Classifiers}.
\newblock IEEE Transactions on Cognitive and Developmental Systems. 2020; p.
  1--10, in press.

\bibitem{Cong2014}
Cong J, Xiao B.
\newblock {Minimizing computation in convolutional neural networks}.
\newblock In: International Conference on Artificial Neural Networks. Springer;
  2014. p. 281--290.

\bibitem{Hartwell2018}
Hartwell A, Kadirkamanathan V, Anderson SR.
\newblock {Compact Deep Neural Networks for Computationally Efficient Gesture
  Classification From Electromyography Signals}.
\newblock In: Proceedings of the 7th IEEE RAS/EMBS International Conference on
  Biomedical Robotics and Biomechatronics; 2018. p. 891--896.

\bibitem{Farrell2008}
Farrell TR.
\newblock {A comparison of the effects of electrode implantation and targeting
  on pattern classification accuracy for prosthesis control}.
\newblock IEEE Transactions on Biomedical Engineering. 2008;55(9):2198--2211.

\bibitem{Young2011}
Young AJ, Hargrove LJ, Kuiken TA.
\newblock {The effects of electrode size and orientation on the sensitivity of
  myoelectric pattern recognition systems to electrode shift}.
\newblock IEEE Transactions on Biomedical Engineering. 2011;58(9):2537--2544.

\bibitem{Hartwell}
Hartwell A, Kadirkamanathan V, Anderson SR.
\newblock {Person-Specific Gesture Set Selection for Optimised Movement
  Classification from EMG Signals}.
\newblock In: Proceedings of the 38th Annual International Conference of the
  IEEE Engineering in Medicine and Biology Society; 2016. p. 880--883.

\bibitem{Zhang2011}
Zhang X, Chen X, Li Y, Lantz V, Wang K, Yang J.
\newblock {A framework for hand gesture recognition based on accelerometer and
  EMG sensors}.
\newblock IEEE Transactions on Systems, Man, and Cybernetics-Part A: Systems
  and Humans. 2011;41(6):1064--1076.

\bibitem{Lu2014}
Lu Z, Chen X, Li Q, Zhang X, Zhou P.
\newblock {A hand gesture recognition framework and wearable gesture-based
  interaction prototype for mobile devices}.
\newblock IEEE transactions on human-machine systems. 2014;44(2):293--299.

\bibitem{Jiang2016}
Jiang S, Lv B, Sheng X, Zhang C, Wang H, Shull PB.
\newblock {Development of a real-time hand gesture recognition wristband based
  on sEMG and IMU sensing}.
\newblock In: Robotics and Biomimetics (ROBIO), 2016 IEEE International
  Conference on. IEEE; 2016. p. 1256--1261.

\bibitem{Krasoulis2017}
Krasoulis A, Kyranou I, Erden MS, Nazarpour K, Vijayakumar S.
\newblock {Improved prosthetic hand control with concurrent use of myoelectric
  and inertial measurements}.
\newblock Journal of neuroengineering and rehabilitation. 2017;14(1):71.

\bibitem{Guo2017}
Guo W, Sheng X, Liu H, Zhu X.
\newblock {Toward an Enhanced Human–Machine Interface for Upper-Limb
  Prosthesis Control With Combined EMG and NIRS Signals}.
\newblock IEEE Transactions on Human-Machine Systems. 2017;47(4):564--575.

\bibitem{Fang2017}
Fang Y, Zhou D, Li K, Liu H.
\newblock {Interface Prostheses With Classifier-Feedback-Based User Training}.
\newblock IEEE Transactions on Biomedical Engineering. 2017;64(11):2575--2583.

\bibitem{Yang2019}
Yang X, Yan J, Liu H.
\newblock {Comparative Analysis of Wearable A-mode Ultrasound and sEMG for
  Muscle-Computer Interface}.
\newblock IEEE Transactions on Biomedical Engineering. 2019; p. 1--9, in press.

\end{thebibliography}

\end{document}